\documentclass[a4paper]{jpconf}
\usepackage{graphicx}
\graphicspath{{Pictures/}}
\begin{document}
\title{\Large Probing the magnitude of asymmetries in the lateral density distribution of electrons in EAS}

\author{Animesh Basak and Rajat K. Dey}

\address{Department of Physics, University of North Bengal, Siliguri, WB 734 013 India}

\ead{rkdey2007phy@rediffmail.com}
\begin{abstract}
The lateral density distributions (LDD) of inclined cosmic ray air shower are asymmetric and the corresponding iso-density contours are of increasing eccentric ellipses with zenith angles of different showers. The polar asymmetry of the iso-density contours introduces a significant shift of the EAS core, which is quantitatively expressed as a gap length (GL) parameter between the EAS core and the center of the modified density pattern consisting of several equi-density ellipses. The LDD of EAS particles is usually approximated by a particular type of lateral density function (LDF) which is generally assumed to be polar symmetric about the EAS axis, and cannot describe the asymmetric LDDs accurately. A polar angle-dependent modified lateral density function of EASs has been derived analytically by considering the effect of attenuation of EAS particles in the atmosphere. From the simulation studies, it has been found that the GL manifests sensitivity to the cosmic ray mass composition. The cosmic ray mass sensitivity of the lateral shower age is also re-examined by applying the modified LDF to the simulated data.
\end{abstract}
\section{Introduction}
Secondary cosmic ray (CR) particles of an extensive air shower (EAS) approaching towards the ground as a thin disk through the atmosphere from the direction of their parent primary CR particle at the speed of light. After the first interaction point, the disk begins to form, and continues to grow, and then starts attenuating after the depth of shower maximum. The transverse and longitudinal momenta imparted on the shower particles emerging from their parent particles via the hadronic interactions would cause the lateral and longitudinal spreads for these particles in an EAS. consequently the periphery of successive iso-density contours get shortened about the EAS axis, which is analogous to an inclined inverted truncated cone.
\section{Cone model: An elliptic lateral density function}
\begin{figure}[htbp]
	\begin{minipage}{18pc}
		\includegraphics[trim=0.8cm 0.8cm 0.8cm 1.5cm, clip=true, totalheight=0.22\textheight, angle=0]{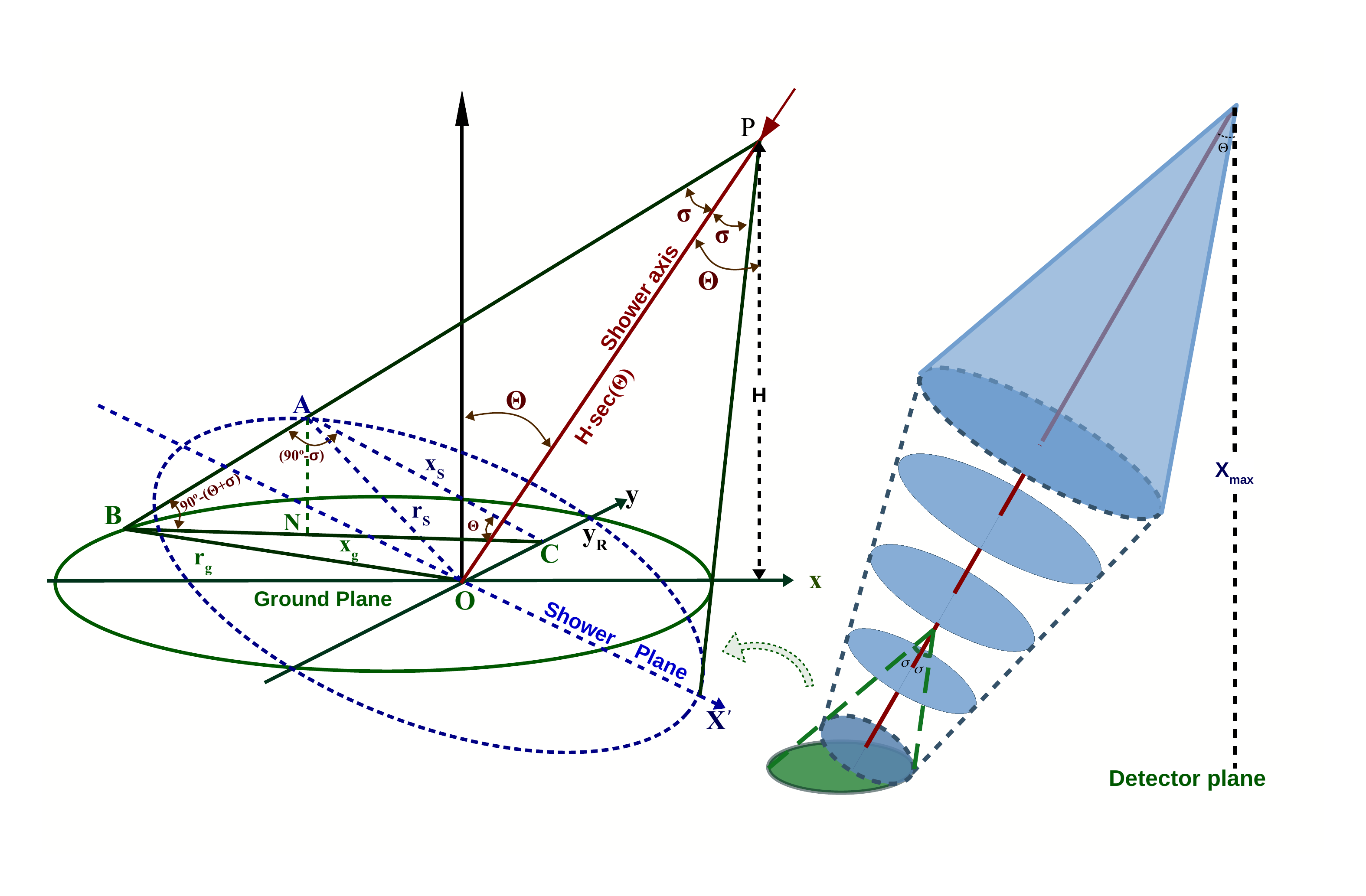}
		\caption{\label{Cone_Prof}Conical shower profile.}
	\end{minipage}\hspace{1pc}%
	\begin{minipage}{18pc}
		\includegraphics[trim=0.6cm -0.2cm 0.6cm -2.2cm, clip=true, width=0.16\textheight]{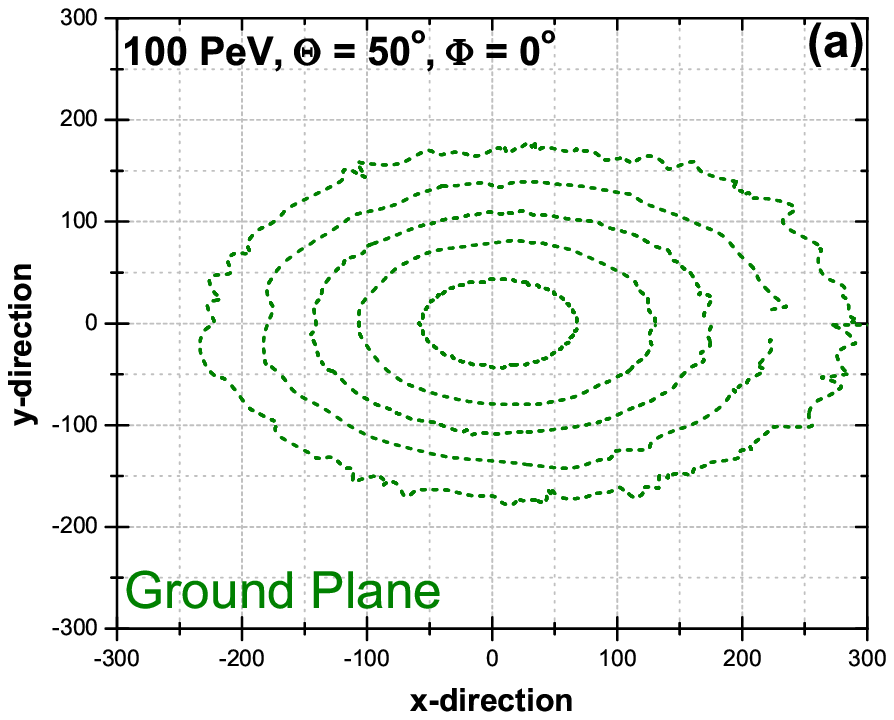}
		\includegraphics[trim=0.6cm -0.2cm 0.6cm -2.2cm, clip=true, width=0.16\textheight]{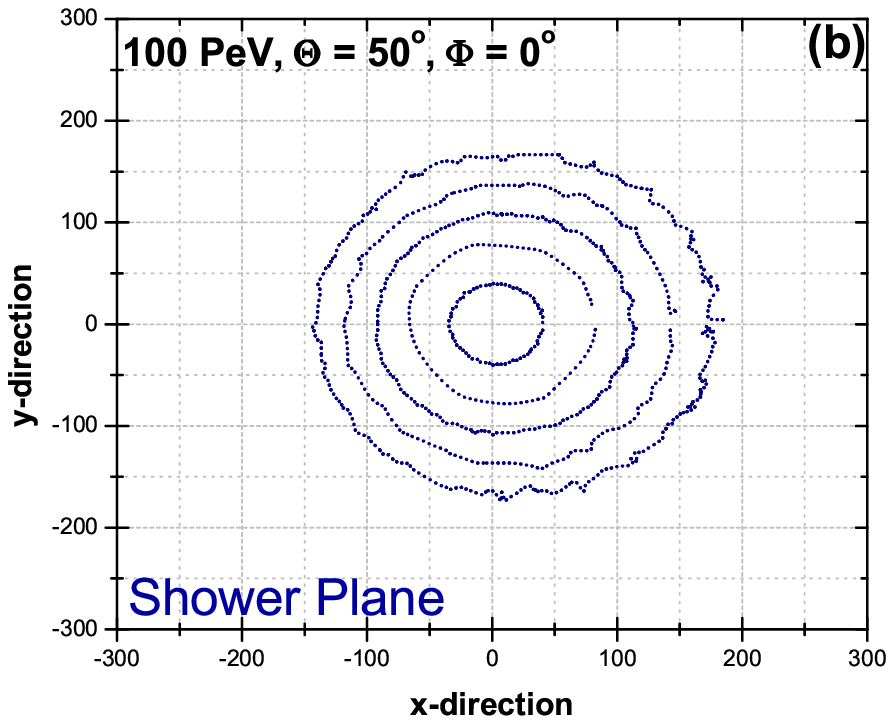}
		\caption{\label{2D_IsoDen} 2-dimensional iso-density contours in the ground and shower plane.}
	\end{minipage}
\end{figure}

The evolution of a conical shower profile of an inclined EAS is shown in the Fig. \ref{Cone_Prof}. The geometric correction is done through the projection of the horizontal elliptic surface to the shower plane. Corresponding equi-density contours are shown in Fig. \ref{2D_IsoDen}(a), (b). The projected electron density in the shower plane ($\rho_s$) is
\begin{equation}
\label{eq:Geom_Corr}
\rho_s(r_s)  =  {\rho_g(r_g)}/{ \cos\Theta}
\end{equation}
An exponential fall of the density of the shower electrons results from the EAS attenuation with a factor $ e ^ {-\eta \cdot AB}$, where $\eta$ is the attenuation length [1]. Electron density in the ground plane $(\rho_g)$ would be 
\begin{equation}
\label{eq:Attn_Dens}
\rho_g(r_g)= \cos\Theta \cdot \rho_s(r_s) \cdot e^{-\eta \cdot AB}
\end{equation}

A characteristic function (CF) to describe the exponential behaviour of LDD with $r_s$ is proposed as follows,
\begin{equation}
\label{eq:CF}
\rho(r_s) \simeq c \cdot e^{-\alpha (\frac{r_s}{r_0})^\kappa}
\end{equation}
Finally the gap length parameter is given by,
\begin{equation}
\label{eq:Xc_1}
x_C = 6813 y_R^{2-\kappa} r_0^\kappa \eta (\alpha \kappa)^{-1}
	\frac{\tan \Theta}{\cos(\Theta+\sigma)} 
	\cdot
	\frac{\cos\sigma}{H-r_s \sin \Theta}
\end{equation}
Since $x_C>0$, the shower attenuation does shift the center of the ellipse towards the early part of the shower. Let us write the above equation as $x_C = 2 A_f y_R \tan\Theta$, where $A_f$ stands for,
\begin{equation}
\label{eq:Af}
A_f = 6813 r_0^\kappa \eta (\alpha \kappa)^{-1}
\cdot
\frac{\cos\sigma}{2\cos(\Theta+\sigma)(H-r_s \sin \Theta)}
\end{equation}
The modified length of semi-minor axis of an equi-density ellipse is,
\begin{equation}
\label{eq:yR}
	y_R = -2 A_f r_g \cos\beta_g \tan\Theta \cdot \frac{\cos^2(\Theta + \sigma)}{\cos^2\sigma}
	+ r_g\sqrt{1-\cos^2\beta_g \sin^2(\Theta+\sigma)}
\end{equation}
We are familiar with most commonly used LDF as NKG function in CR air shower physics.
The polar angle dependent elliptic-LDF (ELDF) can be found from the NKG type Symmetric-LDF (SLDF) by making a substitution for the variable $r_s$ by $y_R$, and the equation for the ELDF finally takes the following structure. 
\begin{equation}
\label{eq:ELDF}
	\rho(r_g,\beta_g)=\cos\Theta\cdot C(s_\perp)N_e \cdot (y_R/r_0)^{s_\perp-2} (1+y_R/r_0)^{s_\perp-4.5}
\end{equation}
Where, $C(s_\perp)=\frac{\Gamma(4.5-s_\perp)}{2\pi r_0^2\Gamma(s_\perp)\Gamma(4.5-2s_\perp)}$ is the normalization factor, while $s_{\perp}$, $r_0$ and $N_e$ respectively are called the age parameter, the moli\`{e}re radius and shower size.
\section{Results and discussions}
\begin{figure}[h]
	\begin{minipage}[b]{2in}
		\includegraphics[trim=0.6cm 0.6cm 1.cm 1.05cm, clip=true, totalheight=0.17\textheight]{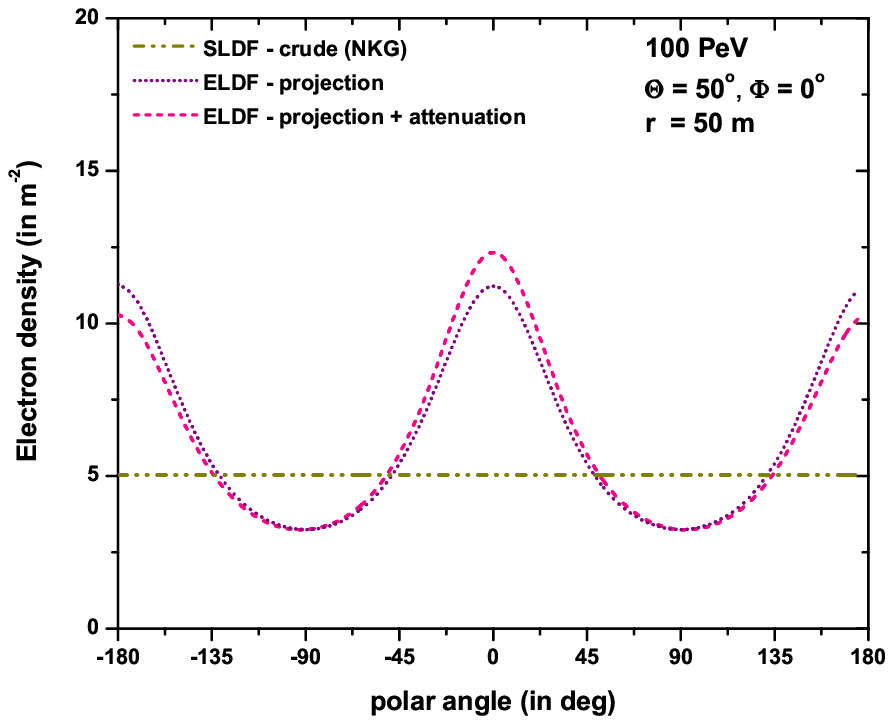}
		\caption{\label{Polar_Dens_ELDF}Ground plane polar density distribution.}
	\end{minipage}\hspace{1pc}%
	\begin{minipage}[b]{2in}
		\includegraphics[trim=0.6cm 0.6cm 0.9cm 0.8cm, clip=true, totalheight=0.17\textheight]{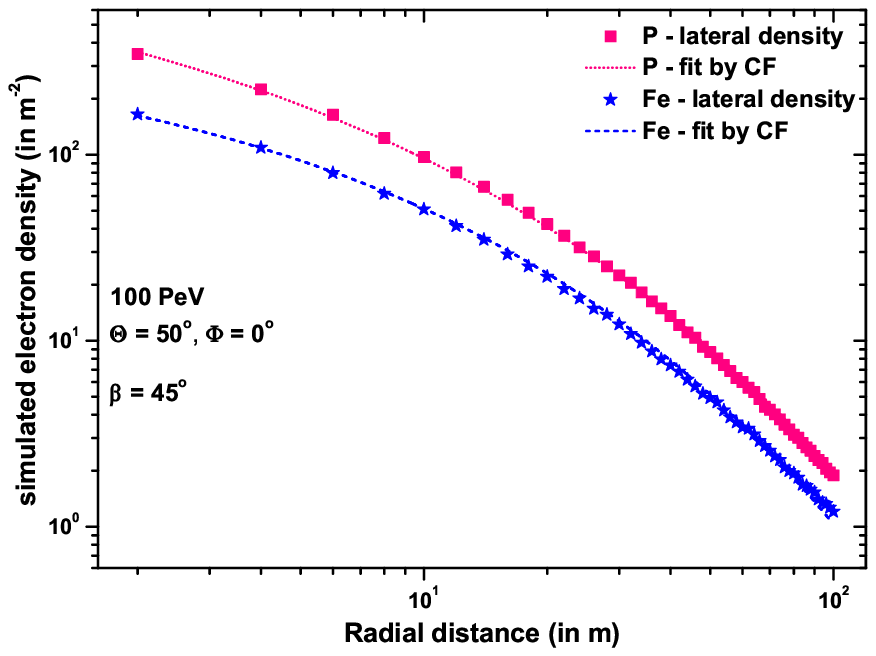}
		\caption{\label{LDD_PI}Electron LDD fitted by CF.}
	\end{minipage}\hspace{1pc}%
	\begin{minipage}[b]{2in}
		\includegraphics[trim=0.6cm 0.6cm 0.6cm 1.05cm, clip=true, totalheight=0.17\textheight]{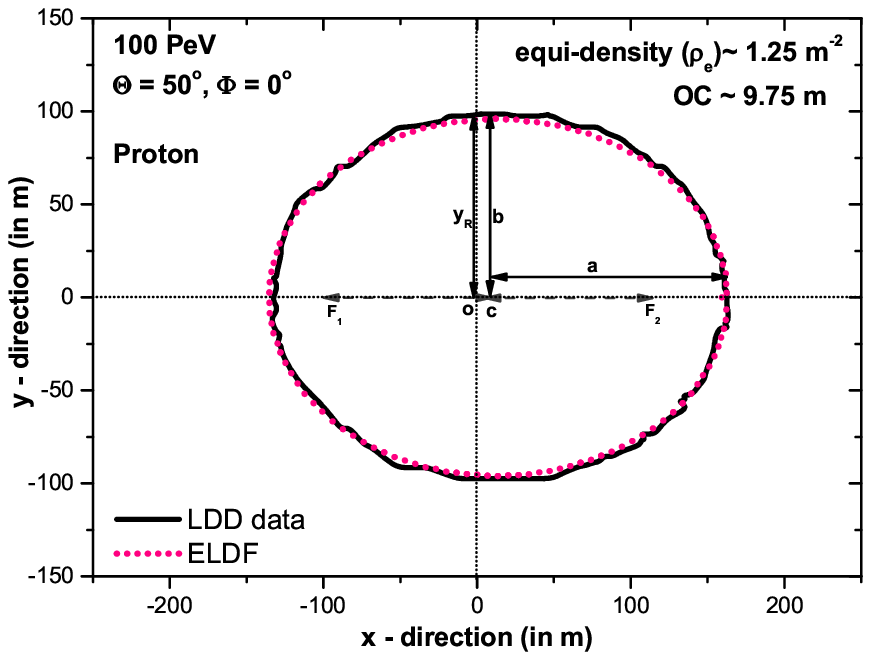}
		\caption{\label{Iso_dens}Formation of GL from equi-density ellipse.}
	\end{minipage}
\end{figure}

\begin{figure}[h]
	\begin{minipage}[b]{2in}
		\includegraphics[trim=0.6cm 0.6cm 0.6cm 1.05cm, clip=true, totalheight=0.17\textheight]{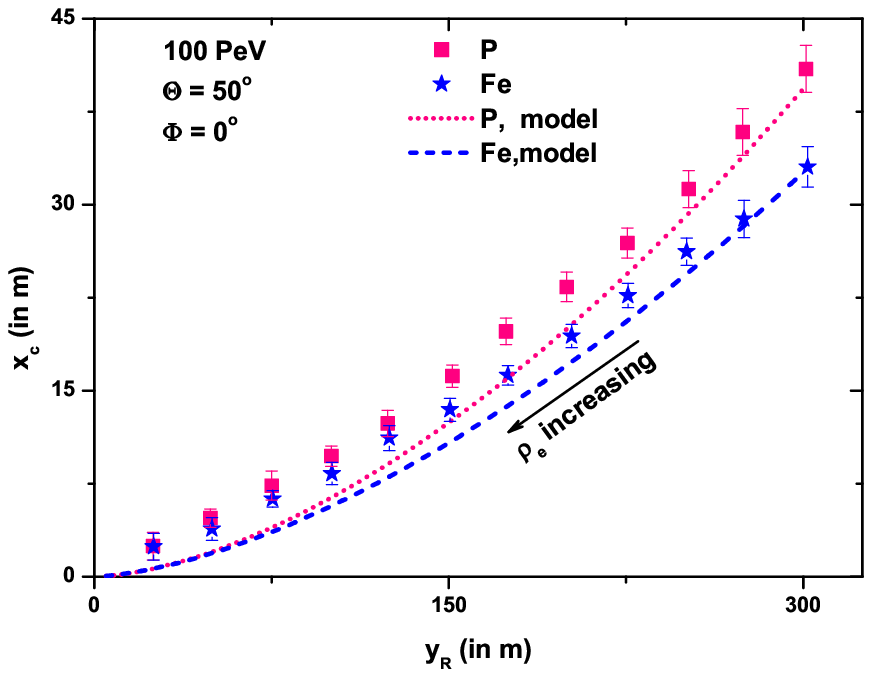}
		\caption{\label{XcYr_PI}$x_C$ vs $y_R$ for P and Fe initiated showers.}
	\end{minipage}\hspace{1pc}%
	\begin{minipage}[b]{2in}
		\includegraphics[trim=0.6cm 0.6cm 0.6cm 1.05cm, clip=true, totalheight=0.17\textheight]{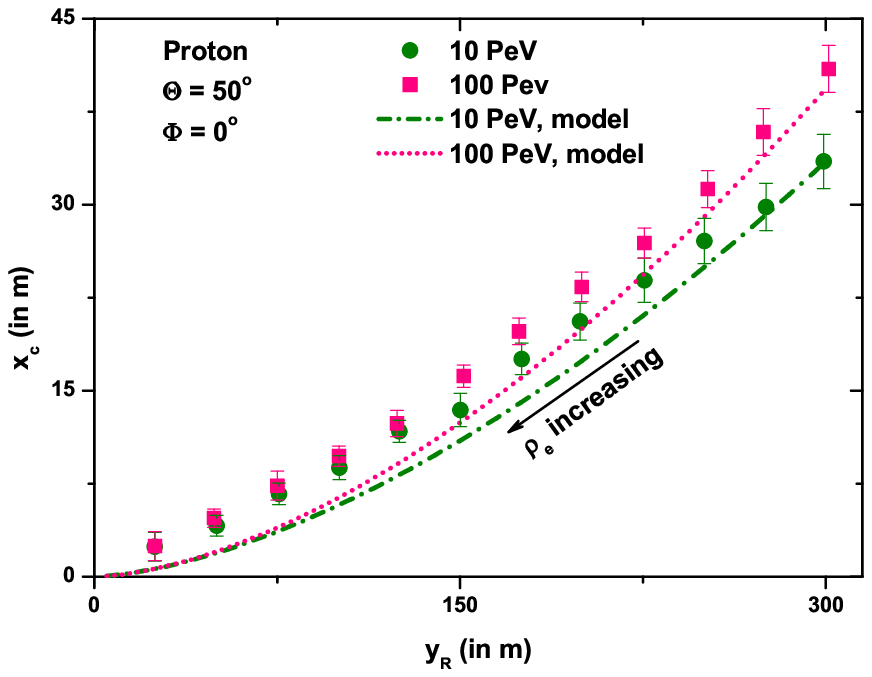}
		\caption{\label{XcYr_E}$x_C$ vs $y_R$ for two energies.}
	\end{minipage}\hspace{1pc}%
	\begin{minipage}[b]{2in}
		\includegraphics[trim=0.6cm 0.6cm 0.6cm 1.05cm, clip=true, totalheight=0.17\textheight]{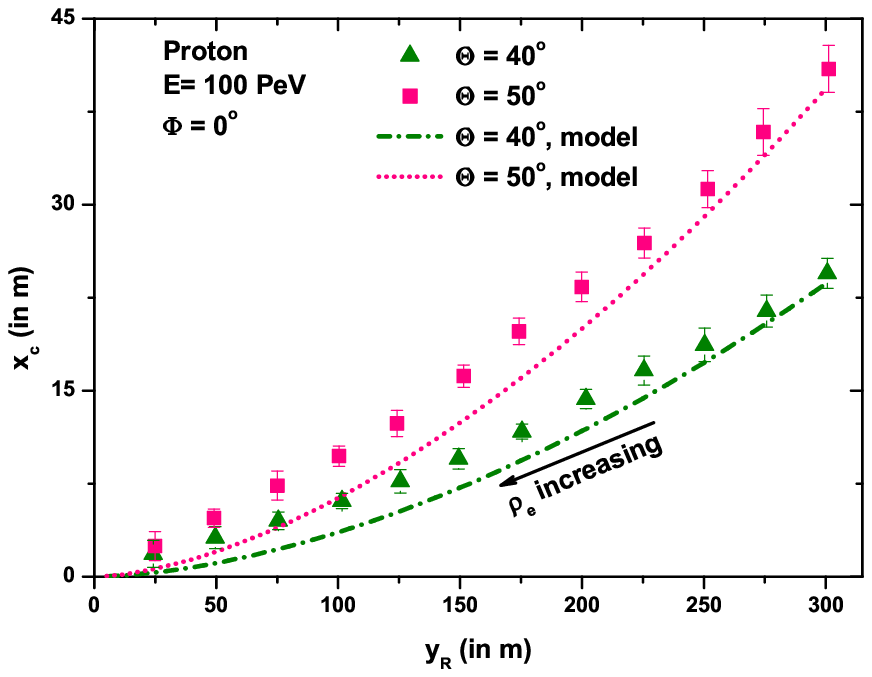}
		\caption{\label{XcYr_Z}$x_C$ vs $y_R$ at two zenith angles.}
	\end{minipage}
\end{figure}

\begin{figure}[h]
	\begin{minipage}[b]{2in}
		\includegraphics[trim=0.6cm 0.6cm 0.6cm 1.05cm, clip=true, totalheight=0.17\textheight]{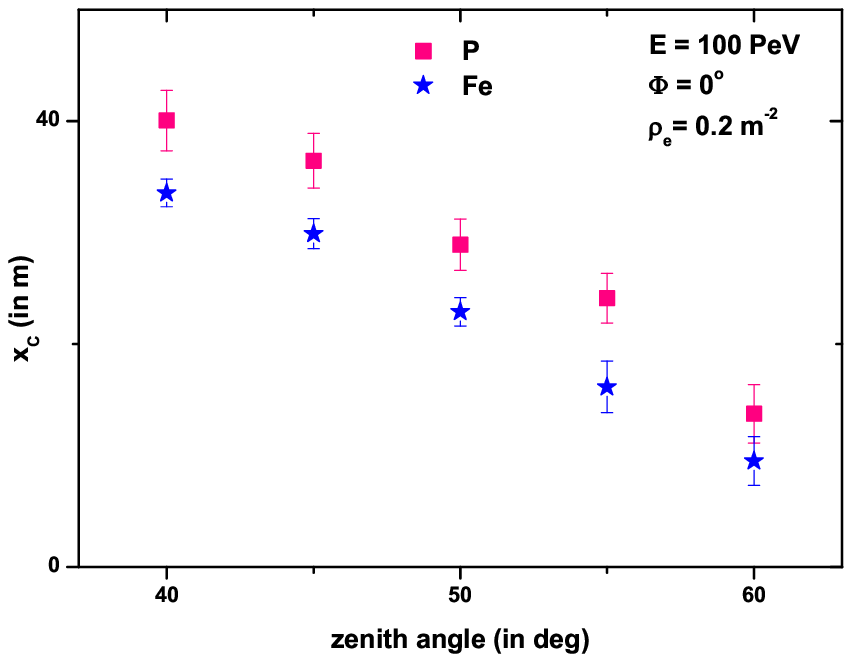}
		\caption{\label{GL_Z_IsoD}Mass sensitivity of GL from its variation with $\Theta$ at fixed $\rho_e$.}
	\end{minipage}\hspace{0.8pc}%
	\begin{minipage}[b]{2in}
		\includegraphics[trim=0.6cm 0.6cm 0.6cm 1.05cm, clip=true, totalheight=0.17\textheight]{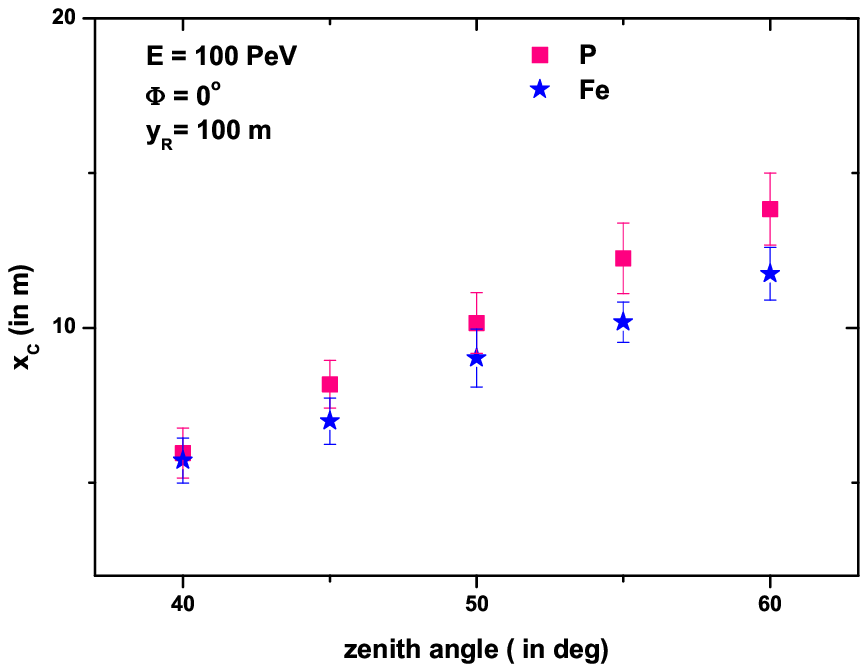}
		\caption{\label{GL_Z_yR}Mass sensitivity of GL from its variation with $\Theta$ at fixed $y_R$.}
	\end{minipage}\hspace{0.8pc}%
	\begin{minipage}[b]{2in}
	\includegraphics[trim=0.6cm 0.6cm 0.6cm 1.05cm, clip=true, totalheight=0.17\textheight]{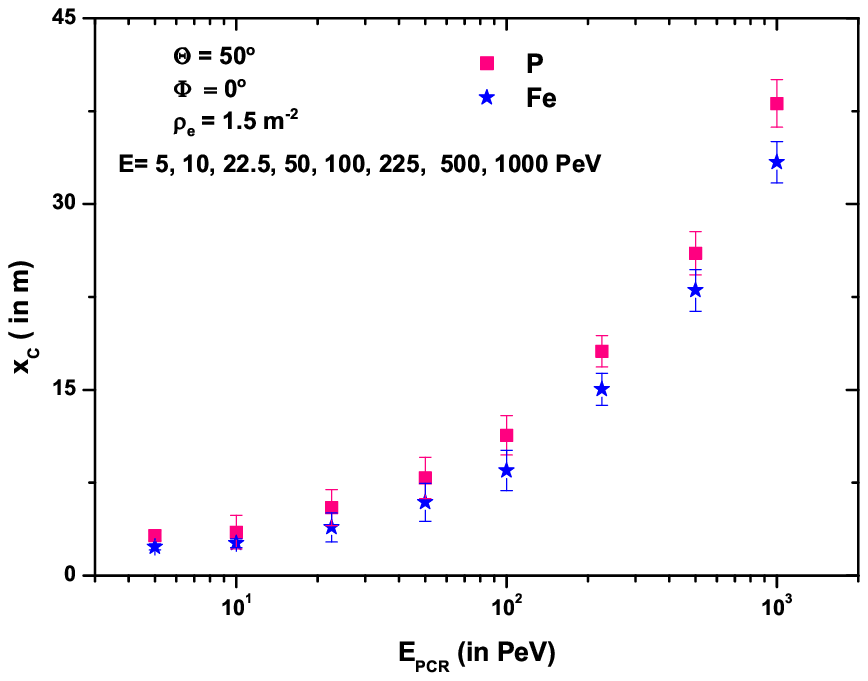}
	\caption{\label{GL_E} Mass sensitivity of GL from its variation with $E$.}
\end{minipage}
\end{figure}

The MC simulation code \textit{CORSIKA} of version 7.69 with the hadronic interaction models QGSJet-01c and UrQMD is used. The reconstructed polar electron densities obtained using SLDF, ELDF including the projection and ELDF including both the attenuation and projection [2], at a core distance 50 m for an average 100 PeV proton shower with $\Theta =50^o$. These are shown in Fig. \ref{Polar_Dens_ELDF}, and the result reconfirms that the ELDF with GL is more appropriate for reconstruction of non-vertical EASs.
In the Fig. \ref{LDD_PI}, the polar averaged LDD for P and Fe initiated showers are approximated by the CF (Eq. \ref{eq:CF}). From the best possible fitting of the simulated data the parameter $\alpha$ picks values 4.3 and 3.7 while $\kappa$ takes 0.36 and 0.43 respectively for P and Fe.
In the Fig. \ref{Iso_dens}, the center of the equi-density ellipse experiences a translation from \textit{O} to \textit{C} $(\overline{OC} \sim 9.75~m)$ solely due to attenuation of EAS electrons. On the other hand, the model predicted GL is about $6.35$~m, evaluated using the Eq. \ref{eq:Xc_1}. 
The GL parameter exhibits sensitivity to P and Fe initiated showers significantly for low-end values of $\rho_e$ (Fig. \ref{XcYr_PI}). GL is found to increase  with energy of CRs (Fig. \ref{XcYr_E}). The elongation of the iso-density curve with increasing $\Theta$ is evident from the values of GL for different zenith angles (Fig. \ref{XcYr_Z}). The model predicted values for GL which are shown by the dotted and short dashed lines (Fig. \ref{XcYr_PI}-\ref{XcYr_Z}) are in good agreement with the simulated data. We have studied the dependence of the GL on $\Theta$ for a fixed electron density (Fig. \ref{GL_Z_IsoD}) as well as at fixed $y_R$ value (Fig. \ref{GL_Z_yR}), which shows a mass sensitivity of primary CR. A correlation between the mean GL with primary energy ($E$) corresponding to $\Theta = 50^o$ and $\rho_e = 1.5~m^{-2}$ for P and Fe induced EASs, is depicted in Fig. \ref{GL_E}.

\begin{figure}[htbp]
	\begin{center}
		\begin{minipage}[b]{2.5in}
			\includegraphics[trim=0.6cm 0.6cm 0.6cm 1.05cm, clip=true, totalheight=0.17\textheight]{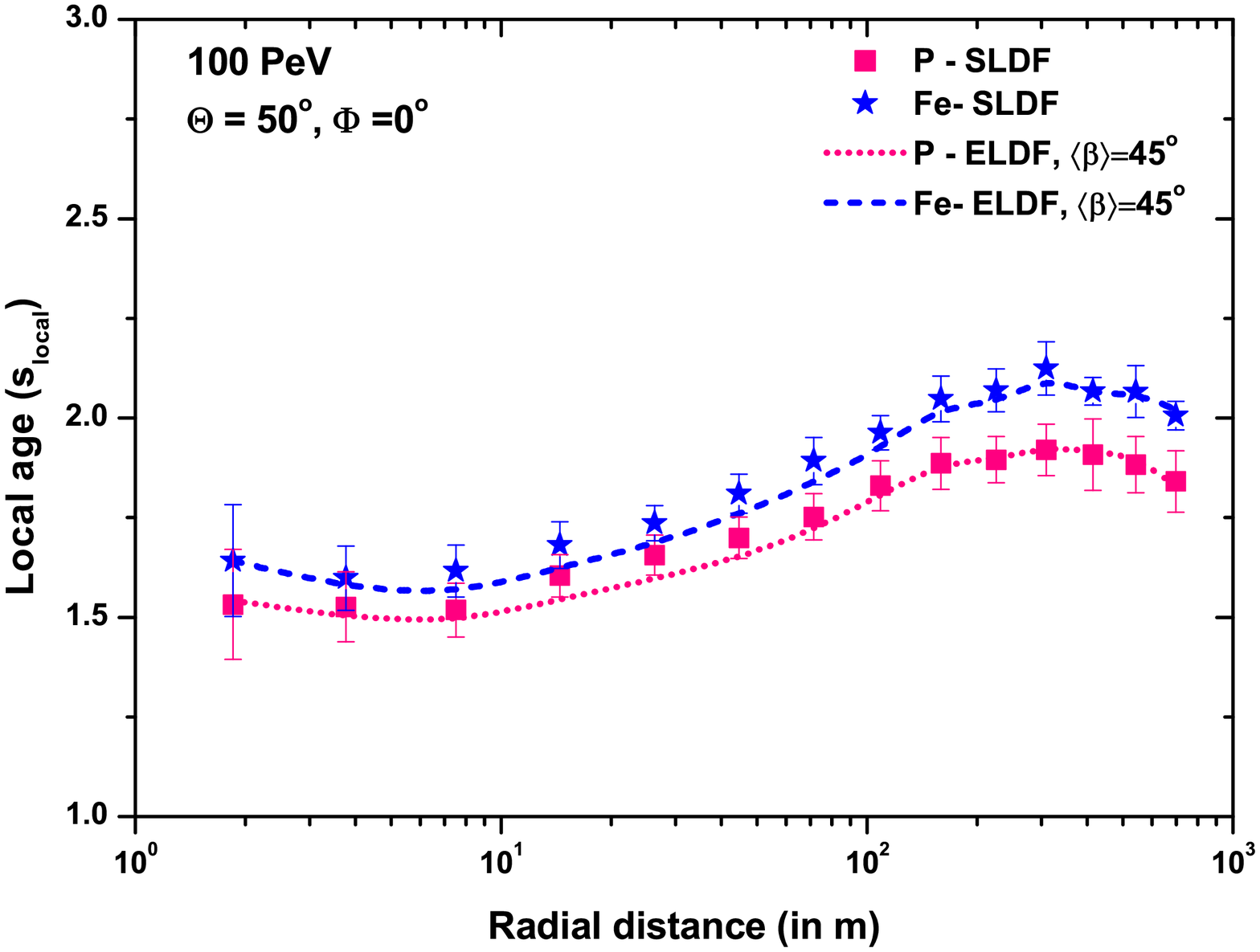}
			\caption{\label{LAP_PI}Variation of LAP with $r_g$.}
		\end{minipage}\hspace{1pc}%
		\begin{minipage}[b]{2.5in}
			\includegraphics[trim=0.6cm 0.6cm 0.6cm 1.05cm, clip=true, totalheight=0.17\textheight]{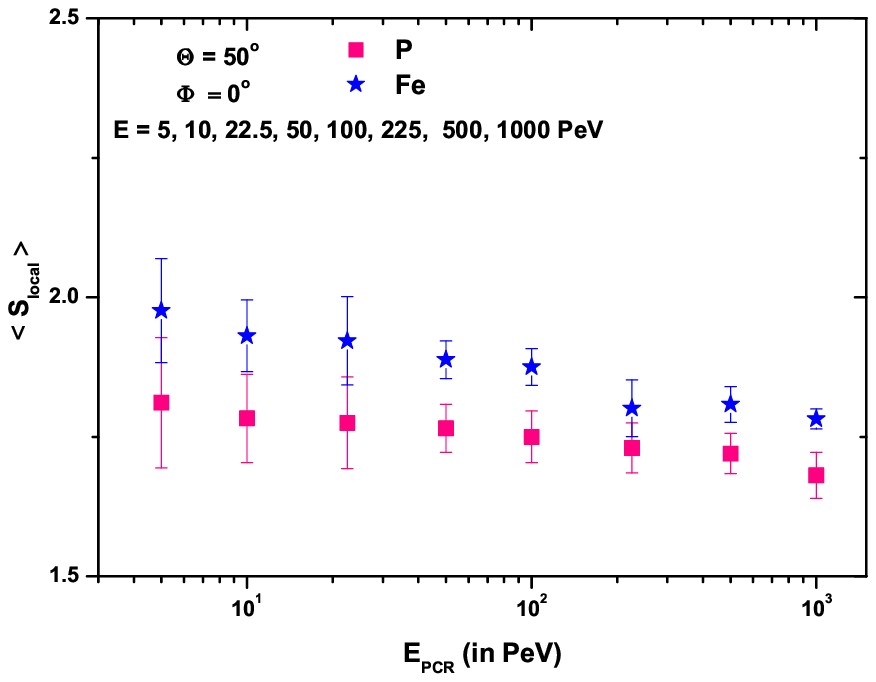}
			\caption{\label{MPAL_PI} Mean LAP versus $E$.}
		\end{minipage}
	\end{center}
\end{figure}

Application of ELDF has been incarnated in terms of the local age parameter (LAP) in Fig. \ref{LAP_PI}.
The analytical expression for the LAP [3] between two adjacent radial distances $[r_i, r_j]$ is:
\begin{equation}
\label{eq:LAP}
s_{ij}=\frac{ln(F_{ij}X_{ij}^2Y_{ij}^{4.5})}{X_{ij}Y_{ij}} 
\end{equation}
Here, $F_{ij}=\rho(y_R(i))/\rho(y_R(j))$,
$X_{ij} = y_R(i)/y_R(j)$, 
$Y_{ij} = (\frac{y_R(i)}{r_0}+1)/(\frac{y_R(j)}{r_0}+1)$
and $r_0$ is the moli\`{e}re radius obtained from the best fit value of LDD by CF. A correlation between the mean LAP with primary energy for P and Fe induced EASs are given in Fig. \ref{MPAL_PI}.
\section{Conclusions \& future outlook}
In this work a modeling of the atmospheric attenuation effect on the LDD of electrons is made considering the conical shower profile. 
The magnitude of the GL that determines the attenuation power of shower particles for a non-vertical EAS, possesses a clear primary CR mass dependence. The ELDF has been used to the simulated electron densities to estimate the LAP, which manifests different radial variation. The variation of mean LAP with primary energy/shower size clearly shows sensitivity to CR mass composition. 

There is a scope to judge the high energy hadronic interaction model dependence of our results in future. An analysis of simulation data considering the LDD of muons and also the combined LDD of electrons and muons are in progress.

\section*{Acknowledgments}
\noindent    
RKD acknowledges the financial support under grant. no. 1513/R-2020. from the University of North Bengal.

\end{document}